\documentclass[11pt]{revtex4-1}

\usepackage{bm}
\usepackage{ulem}
\usepackage{color,graphicx}
\usepackage{tensor}
\usepackage{amsmath}
\usepackage{tikz}
\usetikzlibrary{calc,decorations.markings}
\usetikzlibrary{shapes.misc}
\tikzset{cross/.style={cross out, draw,
         minimum size=2*(#1-\pgflinewidth), inner sep=0pt, outer sep=0pt},cross/.default={3pt}}

\usepackage[colorlinks,pdfusetitle,urlcolor=blue,citecolor=blue,linkcolor=blue,bookmarksnumbered,plainpages=false]{hyperref}

\begin{document}

\title{Theory of paraxial optical Skyrmions}

\author{Z. Ye, S. M. Barnett, S. Franke-Arnold, J. B. Götte,
A. McWilliam, F. C. Speirits,  and C. M. Cisowski}

\email{stephen.barnett@glasgow.ac.uk}
\affiliation{School of Physics and Astronomy, University of Glasgow,
Glasgow G12 8QQ, United Kingdom}

\date{\today}

\begin{abstract}

Vector light beams, characterised by a spatially varying polarisation, can exhibit localised structures 
reminiscent of the Skyrmions familiar from the study of magnetic media.  We present a theory of such Skyrmions within paraxial optics, exploiting mathematical analogies with the study of superfluids, especially the A phase of superfluid ${\rm He}^3$. 
 The key feature is the Skyrmion field which, together with the 
underlying Skyrmion vector potential, determines the properties of the Skyrmions and, more
generally, the polarisation structure of every paraxial vector beam.  In addition to structures with integer Skyrmion number we find polarisation patterns with non-integer Skyrmion number; these seem to have no analogue in other fields of physics.

\end{abstract}

\pacs{}
\maketitle


\section{Introduction}

Paraxial light beams with a spatially varying polarisation pattern underpin many recent developments in the
field of structured light~\cite{Forbes21,Rubinsztein-Dunlop,Zhan,Jinwen}. 
Such structured beams can be intricate, with many spatial modes contributing to a complex polarisation pattern~\cite{Roberta,Galvez,Bauer}. 
At their simplest, however, they may be formed from a mere two distinct spatial modes with orthogonal polarisations.
Fundamental examples are the beams with a radial or an azimuthal polarisation~\cite{otte,Nesterov,Oron}.
The former of these is remarkable for enabling tight focusing and enhanced field strengths in the propagation 
direction~\cite{Quabis00,Dorn03}.

We have shown, in 2020, that the simplest vector beams can exhibit a polarisation pattern that is readily identified with a Skyrmionic structure~\cite{Parax}. Such patterns are characterised by a polarisation at the centre of the beam that is orthogonal to that at the edges, together with a polarisation that rotates on the Poincar\'{e} sphere as we follow an orbit around the beam centre.

The structures 
that came to be known as Skyrmions were first introduced by Skyrme as topological excitations 
in a non-linear field theory~\cite{SkyrmePRSA} and applied to the theory of mesons and baryons~\cite{SkyrmeNucl}.
These ideas have largely been superceeded, but a simpler purely spatial form of Skyrme's excitations, the so-called baby Skyrmions~\cite{Piette}, has had a wider application in many other areas of physics including quantum liquids~\cite{Vollhardt,Volovik,Leggett}, photonic materials~\cite{Mechelen},  
fractional statistics~\cite{Wilczek}, non-linear field theories~\cite{Pisamen} and even in cosmology~\cite{Strings}.  It is perhaps in magnetic media, however, where the idea has had its greatest impact~\cite{Sachdev,Seki,Dennis11,Bogdanov,Rossler,Muhlbauer,Romming,Dupe,Hsu}, where they have been
proposed for use in data storage~\cite{Fert,Foster}.  
In this context, localised Skyrmions 
can be found in the orientation of the magnetisation in magnetic materials.

In optics there has been a wide variety of recent developments, both theoretical and experimental, associated
with Skyrmions in a multitude of guises.  Skyrmionic structures have been prepared in the interference between
evanescent fields~\cite{Tsesses}, in plasmons associated with metamaterials~\cite{Alu} and in a range of other 
structures~\cite{Shi}. They have been identified as features in yet more complex field structures 
\cite{Shen1,Shen2,Shen3,Denz,Denz2,Parmee,Ciso2023}.
It has been shown that optical Skyrmions can be generated in suitable microcavities~\cite{Lin}.  Optical Skyrmions
have also been shown to exist in the strongly focussed regime, beyond the paraxial regime~\cite{Pisanty}. The simple
paraxial Skyrmions proposed in~\cite{Parax} have been prepared and the associated Skyrmion number measured
\cite{Zhu}.

Here we present the theory of propagating Skyrmions within paraxial optics.  We reveal the role played by the
Skyrmionic field in the propagation properties of Skyrmions and, in particular, its role in the stability of these structures noted in~\cite{Nape}. We show that the Skyrmion field is transverse (divergenceless) and, therefore that it can be written as the curl of a further vector field in the same manner that the magnetic induction in 
electromagnetism is the curl of the vector potential.  This idea has been exploited in the theory of superfluid
helium 3~\cite{Vollhardt} and we translate 
concepts from this topic into paraxial optics to 
underscore the value of the Skyrmion numbers for 
Skyrmion beams.
We previously introduced a second method for measuring the skyrmion number experimentally, showing its advantage for optical applications~\cite{mcwilliam}. Here, we derive this method in a general context and highlight specific cases of interest. 
Finally, we investigate the exotic properties of paraxial beams with a non-integer Skyrmion number.


\section{Skyrmions and their mathematical properties}

Our construction of paraxial optical Skyrmions has features in common with the magnetic Skyrmions formed on the 
surface of a suitable material~\cite{Seki} and these 
provide a natural starting point for our analysis.   
Magnetic Skyrmions can be visualised as a covering of a Bloch sphere on which is wrapped the local magnetisation or, more precisely, the direction of the 
local magnetisation, and with every possible direction present at least at one point in space.  
Magnetic Skyrmions are mapped from the stereographic projection of the 3D sphere onto the 2D plane of the magnetic surface~\cite{Ishikawa,Beille,Lebech1,Lebech2}.  At the centre of the Skyrmion, the direction of the 
magnetisation is opposite to that at large distances from the centre.  The magnetisation changes gradually as we
move from one point to an adjacent one. 
The Skyrmion number, $n$, is then the number of rotations of the magnetisation
around the Bloch sphere as we traverse a closed circuit around the centre of the Skyrmion.  
This Skyrmion number has the mathematical form~\cite{Sachdev,Seki}
\begin{equation}
\label{Eq2.1}
n = \frac{1}{4\pi} \int {\bf 
M}\cdot\left(\frac{\partial {\bf M}}{\partial x} \times \frac{\partial {\bf M}}{\partial y}\right) \mathrm{d}x\,\mathrm{d}y \, ,
\end{equation}
where the integration takes place over the magnetic surface, taken here to define the plane $z = 0$.  The quantity ${\bf M}$
corresponds to the local {\it direction} of the magnetisation.  Note that there is no dependence on the strength of the 
magnetisation and that ${\bf M}$ is a vector field with unit magnitude everywhere. 
Hence we can picture the unit vector field ${\bf M}$ as representing, at each point, the Bloch
vector on the corresponding Bloch sphere.

To make the transition into paraxial optics, we replace the magnetic spin direction {\bf M} with
a vector, {\bf S}, formed by the normalised Stokes parameters associated with the local optical polarisation~\cite{Born,Wolf}.  The vector field points in the direction of the local orientation 
of the magnetic spin, but the Stokes vector is oriented in the abstract space of the Poincar\'{e}
sphere. Despite this analogy, it is important to note that unlike magnetic skyrmions, for which the spin structure is determined by energy constraints defined by the film structure, conservation laws, and energy minimization, 
optical skyrmions are only limited by Maxwell's equations and thus offer a versatile platform for the exploration of exotic topological structures~\cite{Ciso2023}. 

The Skyrmion number can be written as the integrated flux of a Skyrmion field across the plane $z = 0$.  
For paraxial optics we shall show that it is a powerful idea,  with implications for the polarisation structure of the beam.
To this end we introduce the Skyrmion field as
\begin{equation}
\label{Eq2.2}
\Sigma_i = \frac{1}{2}\varepsilon_{ijk}\varepsilon_{pqr}S_p\left(\frac{\partial S_q}{\partial x_j}\right)
\left(\frac{\partial S_r}{\partial x_k}\right) \, ,
\end{equation}
where 
$\varepsilon_{ijk}$ and $\varepsilon_{pqr}$ are the alternating (Levi-Civita) symbols~\cite{Paul} and we employ the
Einstein summation convention for repeated indices. In vectorial notation, Eq.~\ref{Eq2.2} becomes:  
\begin{equation}
\label{Eq2.9}
\Sigma_i = \frac{1}{2}\varepsilon_{ijk}{\bf S}\cdot\left(\frac{\partial {\bf S}}{\partial x_j}\times\frac{\partial {\bf S}}{\partial x_k}\right)
\, .
\end{equation}
The Skyrmion field, 
\mbox{\boldmath$\Sigma$}, is the natural counterpart of the Skyrmion current from
Skyrme's original paper~\cite{SkyrmePRSA}.  In terms of this field, our Skyrmion number is simply
\begin{equation}
\label{Eq2.3}
n = \frac{1}{4\pi}\int \mbox{\boldmath$\Sigma$}\cdot \mathrm{d}{\bf A} \, ,
\end{equation}
where the surface of integration, ${\rm d}{\bf A}$, is the $z = 0$ plane.  The form of the Skyrmion field is closely related to the scalar 
triple product and, indeed, it can be written as a combination of determinants.  From this we can infer an important 
symmetry, which is that the Skyrmion field is invariant under a global rotation of the polarisation
on the Poincar\'{e} sphere (the same rotation
applied to every point).  This statement, which is readily confirmed by direct calculation, means that each optical Skyrmion is part of a 
family with differently oriented polarisations but identical Skyrmion fields~\cite{Parax}.

In our study of optical Skyrmions we shall exploit the form of the Skyrmion field throughout space and we examine here 
the general properties of this field.  The first thing to note is that the field is transverse, or divergenceless:
\begin{equation}
\label{Eq2.4}
\mbox{\boldmath$\nabla$}\cdot\mbox{\boldmath$\Sigma$} = 0 \, .
\end{equation}
To prove this we recall that the alternating symbol is antisymmetric under interchange of any two indices and hence
\begin{align}
\label{Eq2.5}
\mbox{\boldmath$\nabla$}\cdot\mbox{\boldmath$\Sigma$} &=
\frac{1}{2}\varepsilon_{ijk}\varepsilon_{pqr}\left(\frac{\partial S_p}{\partial x_i}\right) 
\left(\frac{\partial S_q}{\partial x_j}\right)
\left(\frac{\partial S_r}{\partial x_k}\right)\, ,  \nonumber \\
&= 3 \det\left(\frac{\partial S_p}{\partial x_i}\right) \, .
\end{align}
To show that this quantity is zero, first consider a general point in the field and denote this by ${\bf r}_0$.  We have not
specified coordinate directions and therefore, without loss of generality, we can define the $x$-direction to correspond to
the orientation of ${\bf S}$ at ${\bf r}_0$.  This, together with the fact that ${\bf S}$ is a unit vector, means that 
${\bf S}({\bf r}_0) = \hat{\bf x}$.  Consider a point ${\bf r}$ that is close to ${\bf r}_0$ and recall that the polarisation
varies continuously in space.  This suggests that we can represent the field ${\bf S}$ at ${\bf r}$ by the first two terms
of the Taylor expansion around its value at ${\bf r}_0$:
\begin{equation}
\label{Eq2.6}
{\bf S}({\bf r}) = {\bf S}({\bf r}_0) + \left.\left[({\bf r} - {\bf r}_0)\cdot\mbox{\boldmath$\nabla$}\right]{\bf S}\right|_{{\bf r}_0} \, .
\end{equation}
From this we can evaluate the length of ${\bf S}({\bf r})$ which is required to be unity:
\begin{align}
\label{Eq2.7}
{\bf S}({\bf r})\cdot{\bf S}({\bf r}) &= {\bf S}({\bf r}_0)\cdot{\bf S}({\bf r}_0) 
+ 2 \left.\left[({\bf r} - {\bf r}_0)\cdot\mbox{\boldmath$\nabla$}\right]{\bf S}\right|_{{\bf r}_0}
 + O\left[|{\bf r} - {\bf r}_0|^2\right]\, ,  \nonumber \\
&\approx 1 + \left.2({\bf r} - {\bf r}_0)\cdot\mbox{\boldmath$\nabla$}S_x\right|_{{\bf r}_0} \, .
\end{align}
We can choose ${\bf r}$ such that ${\bf r} - {\bf r}_0$ points in any desired direction and it follows, therefore, that any
first derivative of the $x$-component at ${\bf r}_0$ is zero.  Note that 
$\mbox{\boldmath$\nabla$}\cdot\mbox{\boldmath$\Sigma$} $, as given in Eq.~(\ref{Eq2.5}), contains first derivatives of all 
three Cartesian components of ${\bf S}$ and this
includes the $x$-component of ${\bf S}$ at our point ${\bf r}_0$.  It then follows that 
$\mbox{\boldmath$\nabla$}\cdot\mbox{\boldmath$\Sigma$} = 0$ everywhere (at least as long as the derivative is well-defined).

The transverse nature of {\boldmath$\Sigma$}, which we have just established, suggests that we can write it as the curl of a
further field, by analogy with the vector potential and magnetic induction for which we write 
${\bf B} = \mbox{\boldmath$\nabla$} \times {\bf A}$.  By analogy with the vector potential in electromagnetism, we refer
to this new field as the Skyrmion vector potential, ${\bf V}$:
\begin{equation}
\label{Eq2.8}
\mbox{\boldmath$\Sigma$} = \mbox{\boldmath$\nabla$}\times{\bf V} \, .
\end{equation}
This idea has long been exploited in the 
study of superfluids and specifically of the A phase of helium 3~\cite{Vollhardt,Salomaa,Volovik}. The advantage of turning to the superfluid literature is that an explicit expression, due to Mermin and Ho~\cite{Vollhardt,Mermin},
exists for the form of the ${\bf V}$ field, our Skyrmion potential becomes:
\begin{equation}
\label{Eq2.10}
\text{V}_i = {\bf m}\cdot \frac{\partial}{\partial x_i}{\bf n} \, ,
\end{equation}
where ${\bf m}$ and ${\bf n}$ are {\it any} two orthogonal unit vector fields such that
\begin{equation}
\label{Eq2.11}
{\bf m}\times{\bf n} = {\bf S} \, .
\end{equation}
It is clear, for example, that we can also write the Skyrmion potential in the form
\begin{equation}
\label{Eq2.12}
\text{V}_i = -{\bf n}\cdot \frac{\partial}{\partial x_i}{\bf m} \, ,
\end{equation}
We note that the non-uniqueness of the fields ${\bf m}$ and ${\bf n}$ is reminiscent of the non-uniqueness of the vector
potential in electromagnetism.  To see this we recall that a gauge transformation of the vector potential has the form
\begin{equation}
\label{Eq2.13}
{\bf A} \rightarrow {\bf A} - \mbox{\boldmath$\nabla$}\chi \, .
\end{equation}
This leaves the magnetic induction unchanged whatever the form of the scalar field $\chi$.  We can arrive at a similar
form of gauge transformation for the Skyrmion potential by first noting that we can write our 
Skyrmion potential in the form
\begin{equation}
\label{Eq2.14}
\text{V}_i = \frac{1}{2}\Im \left[({\bf m} - i{\bf n})\cdot \frac{\partial}{\partial x_i}({\bf m} + i{\bf n})\right] \, ,
\end{equation}
where we have used the readily checked properties
\begin{equation}
\label{Eq2.14a}
{\bf m}\cdot\frac{\partial}{\partial x_i}{\bf n} =  0 
= {\bf n}\cdot\frac{\partial}{\partial x_i}{\bf m} \, .
\end{equation}
Our Skyrmion vector potential, Eq. (\ref{Eq2.14}), is clearly unchanged if we replace ${\bf m} + i{\bf n}$ by $e^{-i\chi}({\bf m} + i{\bf n})$ and 
${\bf m} - i{\bf n}$ by $e^{i\chi}({\bf m} - i{\bf n})$ for some constant phase $\chi$, corresponding to a rotation of the
${\bf m}$ and ${\bf n}$ fields about the direction given by the vector ${\bf S}$:
\begin{align}
\label{Eq2.15}
{\bf m} &= \Re({\bf m} + i{\bf n}) \rightarrow {\bf m} \cos\chi + {\bf n} \sin\chi  \nonumber \\
{\bf n} &= \Im({\bf m} + i{\bf n}) \rightarrow {\bf n} \cos\chi - {\bf m} \sin\chi \, .
\end{align}
This is reminiscent of quantum electrodynamics, in which the gauge freedom is related to the non-unique phase 
of the electron wavefunction~\cite{Aitchison}.  If, as in quantum electrodynamics, we allow $\chi$ to be a function
of ${\bf r}$ then the Skyrmion potential becomes 
\begin{equation}
\label{Eq2.16}
{\bf V} \rightarrow {\bf V} - \mbox{\boldmath$\nabla$}\chi \, ,
\end{equation}
precisely as in Eq.~(\ref{Eq2.13}) for the vector potential.  This means that we can choose our vector fields 
${\bf m}$ and ${\bf n}$ as we like at different points in space and change only the Skyrmion potential but not
the Skyrmion field {\boldmath$\Sigma$}.

The relationship between our Skyrmion field and the Skyrmion potential suggests that we might invoke Stokes'
theorem to replace the surface integral over the Skyrmion field by a line integral of the Skyrmion potential. 
There is, however, a subtlety to be accounted for, which is analogous to the theory of 
Dirac strings associated with magnetic monopoles~\cite{Dirac1,Dirac2,Jackson,Lawrie}.  To illustrate this
we choose a specific form for the fields ${\bf m}$ and ${\bf n}$ and, in doing so, a specific form for the Skyrmion
potential.  We start by noting that the ${\bf S}$ field has the form
\begin{equation}
\label{Eq2.17}
{\bf S} = S_x\hat{\bf x} + S_y\hat{\bf y} + S_z\hat{\bf z} \, .
\end{equation}
A natural choice for the ${\bf m}$ and ${\bf n}$ fields is
\begin{eqnarray}
\label{Eq2.18}
{\bf m} &=& \frac{1}{\sqrt{S_x^2+S_y^2}}(S_y\hat{\bf x} - S_x\hat{\bf y})  \nonumber \\
{\bf n} &=&  \frac{1}{\sqrt{S_x^2+S_y^2}}[-S_zS_x\hat{\bf x} - S_zS_y\hat{\bf y} + (S_x^2  + S_y^2)\hat{\bf z}] \, ,
\end{eqnarray}
for which the Skyrmion potential is
\begin{equation}
\label{Eq2.19}
\text{V}_i = \frac{S_z}{S_x^2 + S_y^2}\left(S_y \frac{\partial}{\partial x_i}S_x - S_x \frac{\partial}{\partial x_i} S_y\right) \, .
\end{equation}
The feature that concerns us is any points (or more precisely lines) at which the Skyrmion potential diverges as these 
require special treatment.  At first sight, it would seem that any point at which $S_z = 1$ (and so $S_x = 0 = S_y$) 
is of
this nature, but a careful analysis of the limit as we approach such points shows that ${\bf V}$ is well-behaved and
finite at these locations.  To identify genuine divergences we consider the form of ${\bf V}$ in cylindrical polar coordinates 
with the $z$-axis running through our point of interest.  In this case we can write the azimuthal component of ${\bf V}$
in the form
\begin{equation}
\label{Eq2.20}
\text{V}_\phi = \frac{S_z}{S_x^2 + S_y^2}\frac{1}{\rho}\left(S_y \frac{\partial}{\partial \phi}S_x - S_x \frac{\partial}{\partial \phi} S_y\right) \, .
\end{equation}
Clearly if the bracketed term does not tend to zero as we approach the line by making $\rho \rightarrow 0$ then 
$\text{V}_\phi$ will diverge.  We shall find that divergences of this form are a characteristic of our optical Skyrmions.

\begin{figure}
    \centering
    \begin{tikzpicture}
    \newcommand{\gap}{0.2}
    \newcommand{\bigradius}{3}
    \newcommand{\smallradius}{0.3}
    
    \coordinate (A) at (1,1.5);
    \coordinate (B) at (-1,0.5);
    \coordinate (C) at (-0.5,-1); 
    
    \draw [help lines,->] (-1.25*\bigradius, 0) -- node[at end, below]  {$x$} (1.25*\bigradius,0);
    \draw [help lines,->] (0, -1.25*\bigradius) -- node[at end, left] {$y$} (0, 1.25*\bigradius);
    
    \draw [thick, decoration={ markings,
    mark=at position 0.02 with {\arrow[line width=1.2pt]{>}},
    mark=at position 0.1 with {\arrow[line width=1.2pt]{>}},
    mark=at position 0.2 with {\arrow[line width=1.2pt]{>}},
    mark=at position 0.3 with {\arrow[line width=1.2pt]{>}},
    mark=at position 0.398 with {\arrow[line width=1.2pt]{>}},,
    mark=at position 0.5 with {\arrow[line width=1.2pt]{>}},,
    mark=at position 0.6 with {\arrow[line width=1.2pt]{>}},
    mark=at position 0.693 with {\arrow[line width=1.2pt]{>}},
    mark=at position 0.8 with {\arrow[line width=1.2pt]{>}},
    mark=at position 0.95 with {\arrow[line width=1.2pt]{>}}},
    postaction={decorate}] 
            let \p1 = (A),
                \p2 = (B),
                \p3 = (C),
                \n1 = {atan2(\y1,\x1)},
                \n2 = {atan2(\y2,\x2)},
                \n3 = {atan2(\y3,\x3)},
                \n4 = {asin(\gap/2/\bigradius)},
                \n5 = {asin(\gap/2/\smallradius)}
            in (\n1-\n4:\bigradius)
            -- ($(\p1) + (\n1-\n5:\smallradius)$) arc (\n1-\n5:-360+\n1+\n5:\smallradius)
            -- (\n1+\n4:\bigradius)
            arc (\n1+\n4:\n2-\n4:\bigradius)
            -- ($(\p2) + (\n2-\n5:\smallradius)$) arc (\n2-\n5:-360+\n2+\n5:\smallradius)
            -- (\n2+\n4:\bigradius)
            arc (\n2+\n4:360+\n3-\n4:\bigradius)
            -- ($(\p3) + (\n3-\n5:\smallradius)$) arc (\n3-\n5:-360+\n3+\n5:\smallradius)
            -- (\n3+\n4:\bigradius)
            arc (\n3+\n4:\n1-\n4:\bigradius);

    \node at (-1.2,3.2) {$\mathcal{L}_\infty$};
    \node at (-0.4,0.5) {$\mathcal{L}_1$};
    \node at (-0.3,-0.5) {$\mathcal{L}_2$};
    \node at (0.7,0.9) {$\mathcal{L}_3$};
    
    \node[cross] at (A) {};
    \node[cross] at (B) {};
    \node[cross] at (C) {};
    
    \end{tikzpicture}
    \caption{Line integral of the Skyrmion vector potential, ${\bf V}$, excluding singular points identified by crosses.}
    \label{fig:multiple_singularities}
\end{figure}
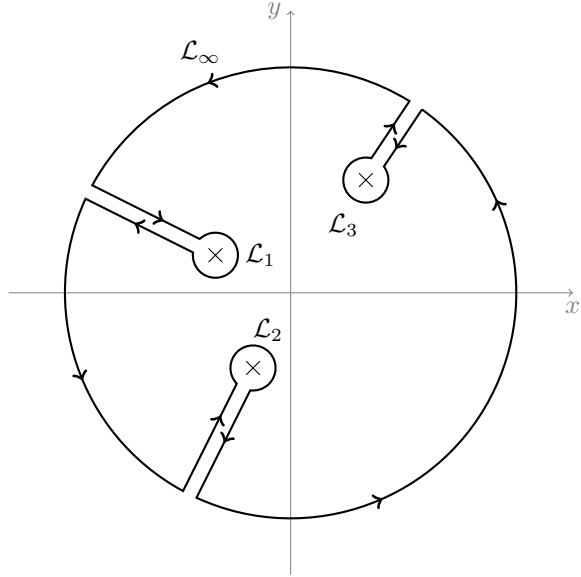

We can use
Stokes's theorem to rewrite our Skyrmion number as a line integral of the Skyrmion potential:
\begin{align}
\label{Eq2.21}
n &= \frac{1}{4\pi}\int \mbox{\boldmath$\Sigma$}\cdot \mathrm{d}{\bf A} \nonumber \\
&= \frac{1}{4\pi}\oint {\bf V}\cdot \mathrm{d} \mbox{\boldmath$\ell$} \, .
\end{align}
As discussed above, it is not sufficient to simply choose a large radius circle centred on the Skyrmion and carry out the integration.
The singular points in ${\bf V}$ must be omitted from the region of integration.  The way in which this is achieved is analogous to that employed in contour integration for dealing with poles in the complex plane~\cite{Paul}.  An example of this is depicted in Fig.~\ref{fig:multiple_singularities}:
the required contour omits any singular points in ${\bf V}$ by passing from the large radius contour, in towards any
singular points, circling them and then returning to the large radius component of the closed contour.
In this way the singularities of ${\bf V}$ are left outside the integration contour.  The line integrals along
the straight lines in from the outer circular path cancel with those in the opposite direction.  
Hence we 
are left only with contributions from the large circular contour and those around and close to the singular
points. Consequently, the Skyrmion number is
\begin{equation}
\label{Eq2.22}
n=\frac{1}{4\pi}\left( \oint_{\mathcal{L}_{\infty}}{\mathbf{V}}\cdot d\mathbf{l}-\sum_{j}{\oint_{\mathcal{L}_j}{\mathbf{V}}}\cdot d\mathbf{l} \right) \, ,
\end{equation}
where the two contours ${\mathcal{L}_{\infty}}$ and $\mathcal{L}_j$ are the circular contour of very large radius and
the inner contour as depicted in Fig.~\ref{fig:multiple_singularities}. 
We shall show in the next section how this can be applied to a paraxial optical Skyrmion, and how we can use winding numbers to obtain a new topological expression for the Skyrmion number.

%


\section{Paraxial optical Skyrmions}

We can construct Skyrmions in paraxial optics by close analogy with those for the magnetisation of suitable surfaces. For magnetic skyrmions 
the direction of the magnetic spin is associated with a point on the Bloch sphere~\cite{QIbook},
whereas for paraxial optics we can map the local polarisation onto the Poincar\'{e} sphere~\cite{Born,Wolf} on which each point is associated 
with a polarisation: the North and South poles correspond to circular polarisations, the equator to linear polarisations, 
with all other points associated with elliptical polarisations.  The mapping between the Bloch and Poincar\'{e} spheres
is a satisfactory one, but there is an important physical difference: a point on the Bloch sphere corresponds to the same
orientation of the spin but the polarisation on the Poincar\'{e} sphere sits in an abstract space and the points on its surface 
do not correspond to physical directions.

To complete the mapping from magnetisation to optical polarisation we need to select a representation of the vector
${\bf S}$.  The natural way to do this is to employ the description of polarisation in terms of the Jones
vectors~\cite{Hecht}.  It is convenient, but not essential, to use the bra-ket notation from quantum mechanics and to
write the Jones vector as a complex superposition of two orthogonal polarisations, which we denote by $|0\rangle$
and $|1\rangle$.  There is no need to decide at this stage which orthogonal polarisations these represent (they could
be left- or right-circular polarisations, horizontal and vertical or any other orthogonal pair).  In terms of these arbitrary 
polarisations we can define the Pauli operators as
\begin{align}
\label{Eq3.0}
\sigma_x &= |0\rangle\langle 1| + |1\rangle\langle 0| \, , \nonumber \\
\sigma_y &= i(|1\rangle\langle 0| - |0\rangle\langle 1|)\, , \nonumber \\
\sigma_z &= |0\rangle\langle 0| - |1\rangle\langle 1| \, .	     
\end{align}
Let us reiterate that the kets $|0\rangle$ and $|1\rangle$ do not correspond to any specific form of polarisation
and so we are dealing with a rotated Poincar\'{e} sphere.  This is a consequence of the fact that the Skyrmion 
field and Skyrmion number are independent of the orientation of the sphere.

To construct our Skyrmionic optical beams we introduce a pair of spatially varying amplitudes, $u_0({\bf r})$ 
and $u_1({\bf r})$, and associate the polarisation at position ${\bf r}$ in the beam with the ket 
\begin{equation}
\label{Eq3.1}
|\psi({\bf r})\rangle = \alpha u_0({\bf r})|0\rangle + \beta u_1({\bf r})|1\rangle \, ,
\end{equation}
where $\alpha$ and $\beta$ are a pair of complex constants.  For any beam we can choose the orthogonal polarisations,
$|0\rangle$ and $|1\rangle$ in such a way that the two mode functions $u_0({\bf r})$ and $u_1({\bf r})$ are orthonormal.
If the beam is propagating in the $z$-direction, this means that for any value of $z$
\begin{equation}
\label{Eq3.2}
\int u_i^*({\bf r}) u_j({\bf r})\,\mathrm{d}x\,\mathrm{d}y = \delta_{ij} \, .
\end{equation}
That a construction of the form in Eq.~(\ref{Eq3.1}) is always possible is a consequence of the Schmidt deomposition
\cite{Schmidt} familiar from the study of entangled states in quantum theory~\cite{QIbook}.

The advantage of employing the bra-ket notation is that we can use some of the mathematical language of 
quantum theory, which serves both to emphasise the analogy with magnetic Skyrmions and also with 
quantum entanglement~\cite{Spreeuw,otte}. We start by normalising $|\psi({\bf r})\rangle$ to write
\begin{equation}
\label{Eq3.3}
|\psi({\bf r})\rangle = \frac{|0\rangle + \mu({\bf r})|1\rangle}{\sqrt{1 + |\mu({\bf r})|^2}} \, ,
\end{equation}
where $\mu({\bf r}) = \beta u_1({\bf r})/\alpha u_0({\bf r})$.  The local polarisation, or more precisely the local
direction on the Poincar\'{e} sphere, is then simply the expectation value of a vector operator the components 
of which are the familiar Pauli matrices:
\begin{equation}
\label{Eq3.4}
{\bf S} = \langle\psi({\bf r})|\mbox{\boldmath$\sigma$}|\psi({\bf r})\rangle \, .
\end{equation}
Hence the components of ${\bf S}$ are simply the normalised Stokes parameters for the beam 
\cite{Born,Wolf,Hecht}.

It is straightforward to determine the components of ${\bf S}$ from $|\psi({\bf r})\rangle$ given in Eq.~(\ref{Eq3.3}):
\begin{align}
\label{Eq3.5}
S_x &= \frac{2\Re(\mu)}{1 + |\mu|^2} \, ,    \nonumber \\
S_y &= \frac{2\Im(\mu)}{1 + |\mu|^2} \, ,    \nonumber \\
S_z &= \frac{1 - |\mu|^2}{1 + |\mu|^2}  \, ,
\end{align}
where we have used the representation of the Pauli operators in Eq.~(\ref{Eq3.0}).

We can use these components to calculate the Skyrmion field by substituting these into Eq.~(\ref{Eq2.9}).  We find a simple,
general expression for the components of the Skyrmion field:
\begin{align}
\label{Eq3.7}
\Sigma_l &= \frac{4\varepsilon_{lmn}}{(1+|\mu|^2)^2}\frac{\partial \Re(\mu)}{\partial x_m}\frac{\partial \Im(\mu)}{\partial x_n} \, , \nonumber \\
&= \frac{\varepsilon_{lmn}}{(1+|\mu|^2)^2}\left(i\frac{\partial \mu}{\partial x_m}\frac{\partial \mu^*}{\partial x_n}
- i\frac{\partial \mu^*}{\partial x_m}\frac{\partial \mu}{\partial x_n}\right)  \, .
\end{align}
By introducing a pair of complex Stokes parameters as
\begin{equation}
\label{EqCS}
S_\pm = S_x \pm iS_y = |S_\pm|e^{\pm i\Phi}, 
\end{equation}
where we have now obtained a phase term, $\Phi$, for this complex quantity, 
the Skyrmion potential has a yet simpler form:
\begin{equation}
\label{Eq3.8}
{\bf V} = -S_z \mbox{\boldmath$\nabla$}\Phi = -S_z \mbox{\boldmath$\nabla$}\arg(\mu) ,
\end{equation}
which is reminiscent of the form of the superfluid velocity in helium 4, which also depends on the gradient of 
the argument of a complex wavefunction~\cite{Tilley,Donnelly}.
Here, however, it is a phase appearing in the superposition of polarisations rather than the argument of a macroscopic wavefunction.  The appearance of the
particular component $S_z$ may seem unexpected, but this component is singled out by being the polarisation corresponding to the
Schmidt basis in $|\psi({\bf r})\rangle$.  It is certainly possible to choose alternative forms for ${\bf V}$,
however, and these are a consequence of the non-uniqueness of the Skyrmion vector potential.

\begin{figure}[htbp]
    \centering
    a.
    \includegraphics[width=5cm]{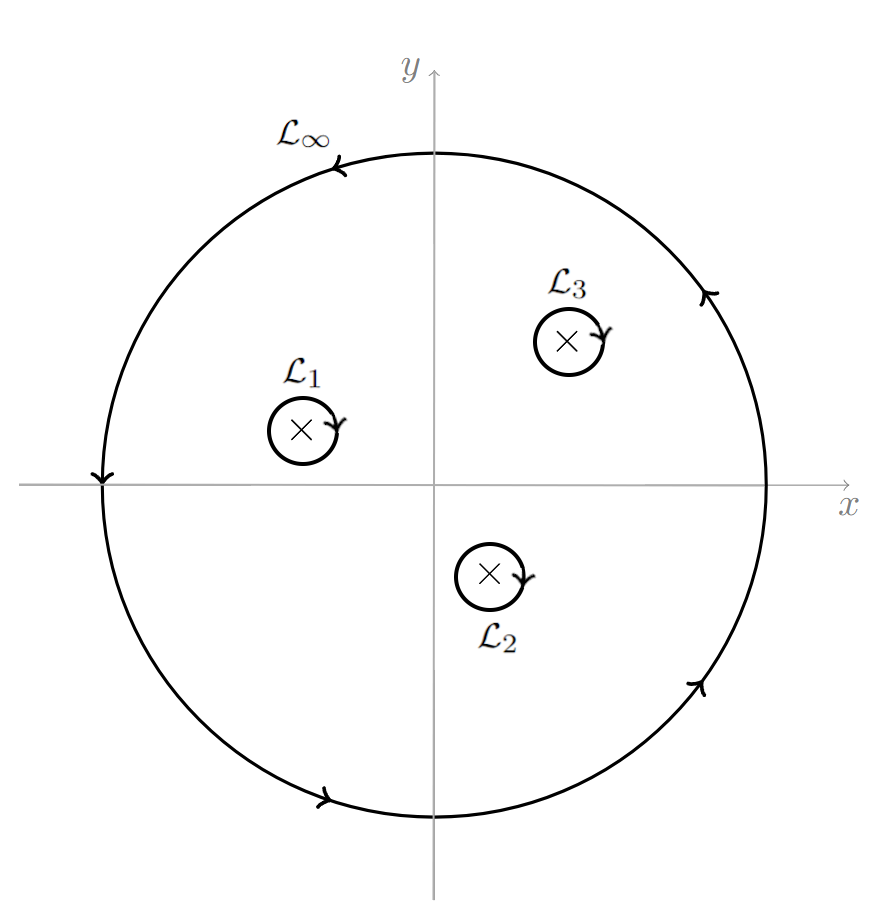}
    b.
    \includegraphics[width=5cm]{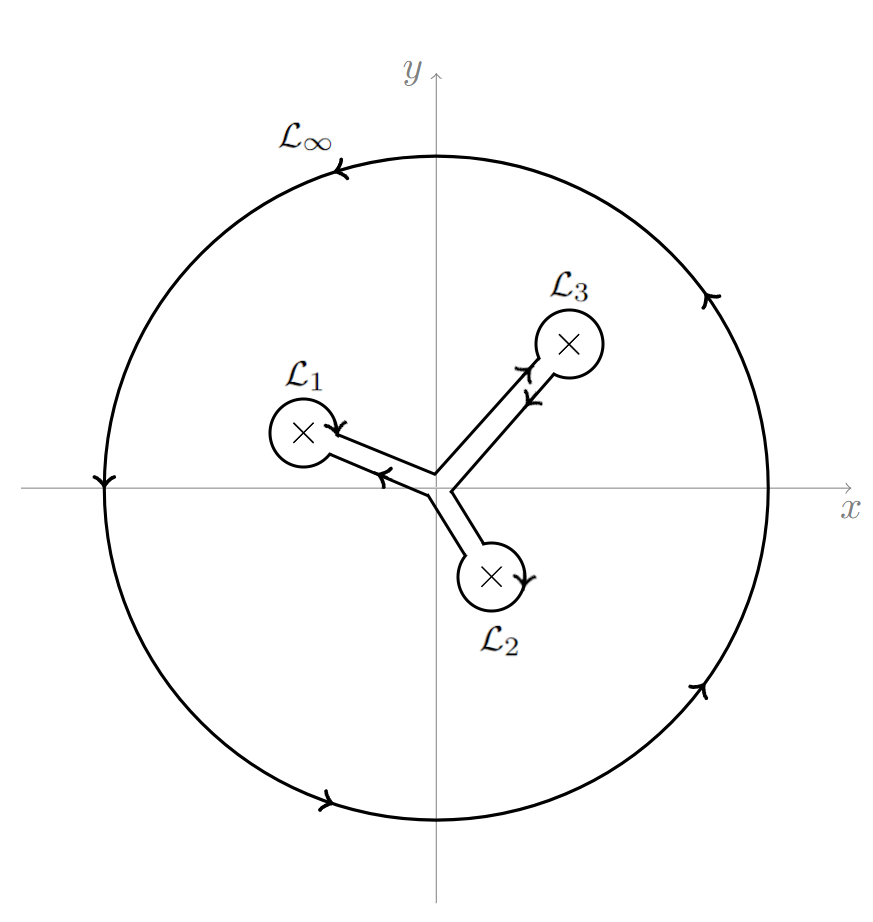}\\
    c.
    \includegraphics[width=5cm]{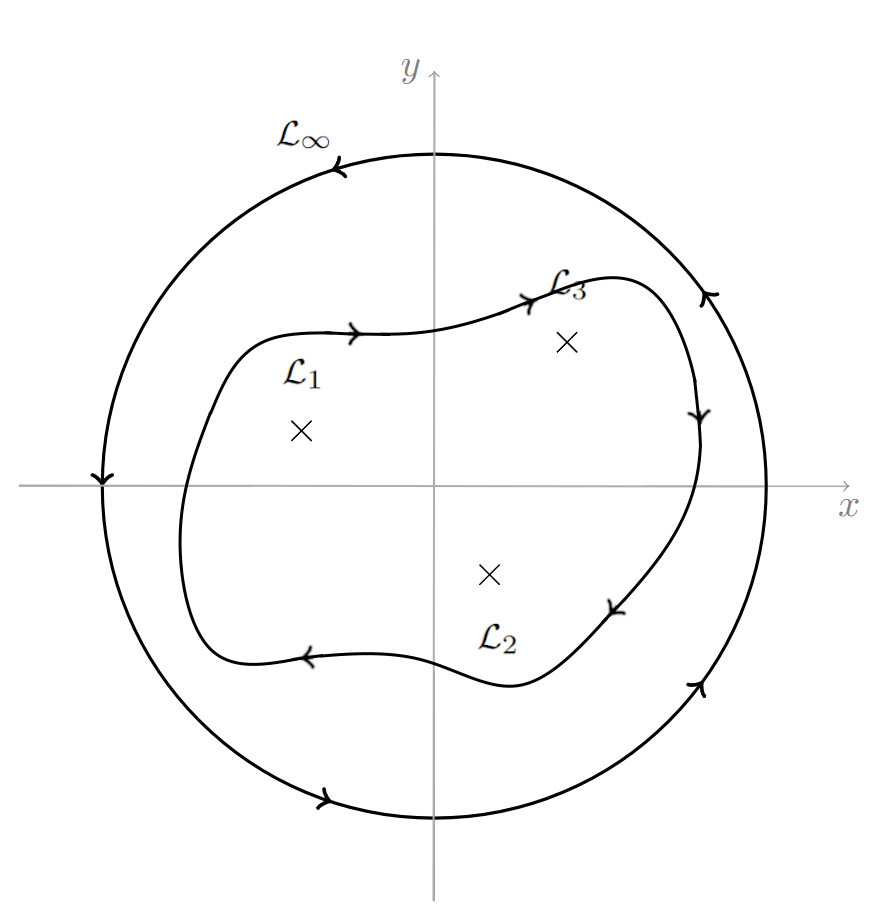}
    d.
    \includegraphics[width=5cm]{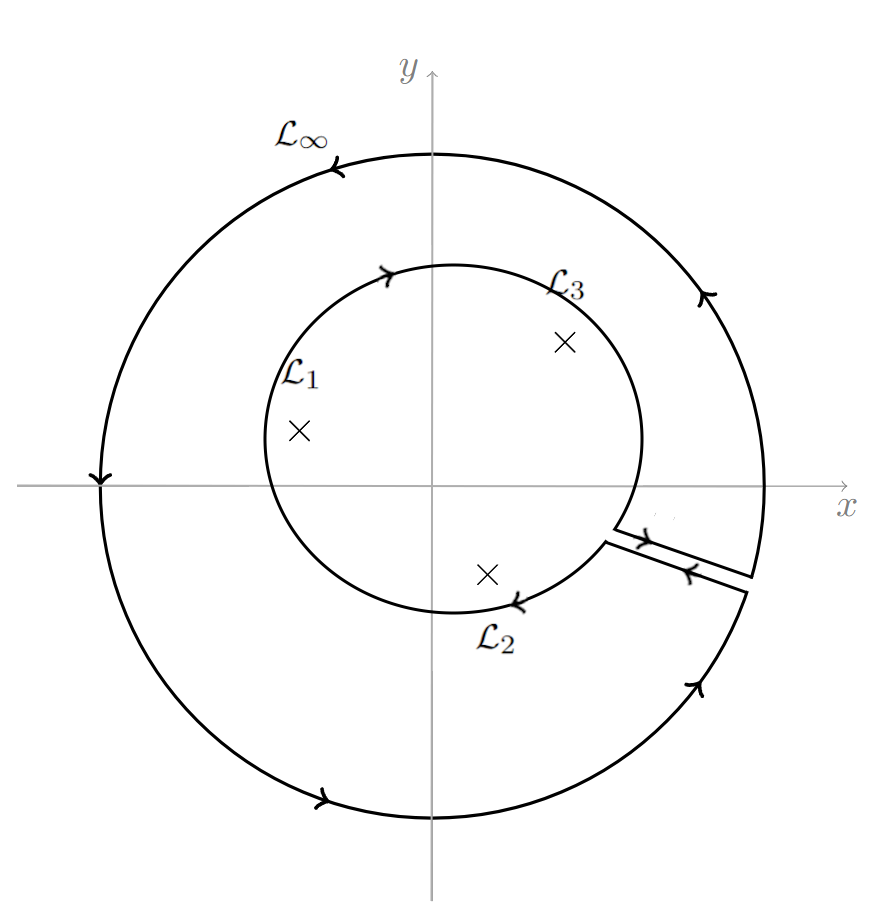}
    \caption{Inner and outer integration contours can be deformed  and stretched (from a to d) without changing the value of the
    Skyrmion number.  
    }
    \label{fig:multiple_singularities2}
\end{figure}

The Skyrmion number defined in Eq.~\ref{Eq2.22} associated with the contour in~Fig.~\ref{fig:multiple_singularities} 
can be written as:
\begin{equation}
    n= \frac{1}{4\pi}\left( \sum_j\oint_{\mathcal{L}_j} S_z\mbox{\boldmath$\nabla$} \Phi \cdot d\bm{l} -\oint_{\mathcal{L}_{\infty}} S_z\mbox{\boldmath$\nabla$} \Phi \cdot d\bm{l}\right),\label{Eq.vIntegral3}
\end{equation}
where $\mathcal{L}_{\infty}$ is the large circular contour and the $\mathcal{L}_j$ are the small circular contours around each of the 
singularities of ${\bf V}$.
Recalling the formula of the winding number $N=(2\pi)^{-1}\oint \mbox{\boldmath$\nabla$} \Phi \cdot d\bm{l}$, we obtain a topological expression for the Skyrmion number:
\begin{equation}
    n = \frac{1}{2}\left( \sum_{j} S_{z}^{(j)} N_j -\bar{S}_{z}^{(\infty)} N_\infty\right),
    \label{Eq.Newskyrmion}
\end{equation}
where $N_\infty$ and $N_j$ are the number of rotations of Stokes vector encountered on circumnavigating, respectively,
the large circular contour and the small contours around each of the singularities.
In spite of the presence of the $1/2$ coefficient, $n$ is guaranteed to be an integer. Indeed, as shown in Fig.~\ref{fig:multiple_singularities2}, the inner integration line around each of the sigularities can always be deformed and stretched to spatial infinity without a change of topological structure such that $ \sum_{j}N_j =N_\infty$ so that:  
\begin{equation}
    n = \frac{1}{2}\sum_{j} (S_{z}^{(j)} -\bar{S}_{z}^{(\infty)}) N_j,
    \label{Eq.Neweq}
\end{equation}
The quantity $\bar{S}_{z}^{(\infty)}$ is the average of $S_z$ obtained by integration around the large circular 
contour.  In the limit that the radius of the contour tends to infinity this should be a constant, but in experimental
realisations an averaging procedure at finite radius is necessary~\cite{mcwilliam}.

The Skyrmion number for any given position along the beam is simply the integral of the $z$-component of this field over the transverse
plane, and so is a function of the $z$-coordinate:
\begin{equation}
\label{Eq3.9}
n(z) = \frac{1}{4\pi}\int \Sigma_z\,\mathrm{d}x\,\mathrm{d}y \, .
\end{equation}
This quantity is remarkably robust in many cases owing to the fact that $\bm{\nabla}\cdot\bm{\Sigma} = 0$,
entailing that the integral of $\bm{\Sigma}$ over any closed surface is zero.  This is a reflection of the general robustness of
structured light beams~\cite{Nape2}. 
It is possible to change the Skyrmion number on propagation, however, as we shall see~\cite{Parax}.


\section{Skyrmion field lines}

The mathematical properties of the Skyrmion field lines are derived directly from their definition.  Firstly, the 
Skyrmion field is transverse, $\mbox{\boldmath$\nabla$}\cdot\mbox{\boldmath$\Sigma$} = 0$, and it follows that the Skyrmion field lines are unbounded; they can exist only as closed loops or extend to infinity.  It follows,
also, that they cannot merge or split, although we shall encounter an interesting exception to this in section
\ref{Sect.fractional}.  Secondly, the Skyrmion field lines are basis independent in that they are unchanged by a
global rotation of the Stokes vector on the Poincar\'{e} sphere.  This means that knowledge of the Skyrmion field
does not determine the polarisation pattern.  It remains, however, to determine the physical significance of the
Skyrmion field lines and we address this here.

Inspection of numerous polarisation patterns leads to the conjecture that Skyrmion field lines are lines of
constant polarisation.  That this is indeed the case was proven in~\cite{Steve2023} but, for completeness, we 
give a summary of the main points here.  We start with the observation that our general polarisation pattern
can be written in the form of a spatially varying ket as given in Eq. (\ref{Eq3.3}).  It follows that the polarisation 
at any given point is determined solely by, and in one-to-one correspondence with, the complex field $\mu({\bf r})$.
Hence lines of constant polarisation are contours of constant $\mu$.

The transverse nature of the Skyrmion field means that at any given point, ${\bf r}_0$, only a single Skyrmion field
line is present and, moreover, that this line is continuous at this point.  It follows that at ${\bf r}_0$ there is a direction 
$u({\bf r}_0)$ along which the Stokes parameters, ${\bf S}$, do not change:
\begin{equation}
\label{Eqskyrmionline}
{\bf u}({\bf r}_0)\cdot\nabla S_i({\bf r}_0)=0 \, .
\end{equation}
As the Stokes parameters do not change in this direction it follows, necessarily, that the polarisation also
does not change.  Hence ${\bf u}({\bf r}_0)$ determines the direction of a line of constant polarisation, which includes but is not restricted to C lines and L lines 
(lines of circular and specific linear polarisation). \cite{Nyebook,Nye,Berry,Dennis_Pi}  

\begin{figure}[htbp]
    \centering
    a.
    \includegraphics[width=7.0cm]{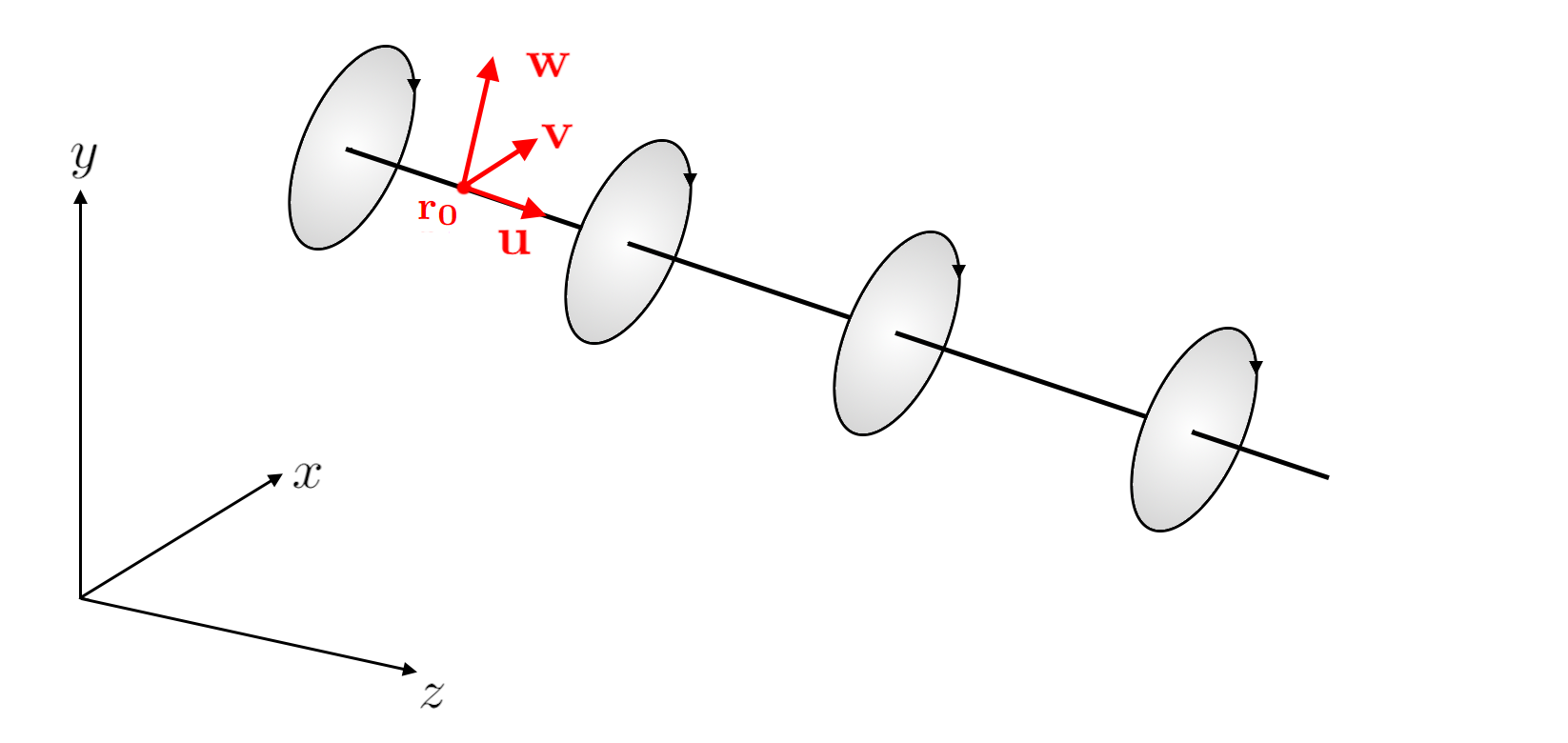}
    b.
    \includegraphics[width=7.0cm]{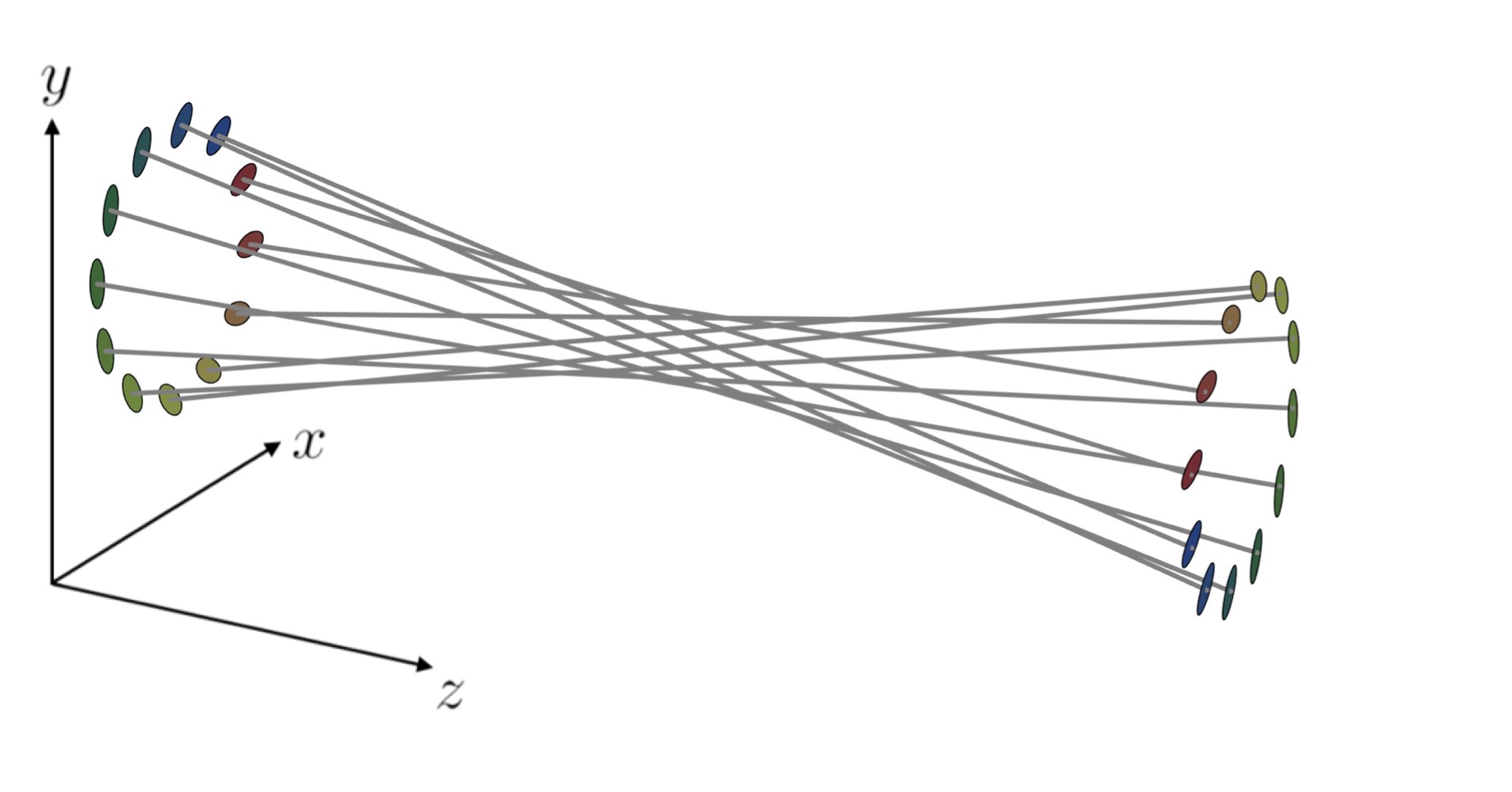}
    \caption{ a. A line of constant elliptical polarisation and the local coordinate system $\bf u$,$\bf v$,$\bf w$ at $\bf{r}_0$. b. A twisted mesh of lines of constant polarisation for a paraxial skyrmion beam ($n=1$) freely propagating over eight Rayleigh ranges.
    }
    \label{fig:line}
\end{figure}

To prove that lines of constant polarisation are also Skyrmion field lines, we return to the definition of the Skyrmion field, given in Eq.~(\ref{Eq2.2}), and evaluate this expression by introducing, at each point, two unit vectors 
${\bf v}({\bf r}_0)$ and ${\bf w}({\bf r}_0)$, where ${\bf u}$,  ${\bf v}$ and ${\bf w}$ are orthonormal vectors 
satisfying the right-hand rule so that, for example, ${\bf u} = {\bf v}\times{\bf w}$, as depicted in 
Fig.~\ref{fig:line} a.  Hence the components of the Skyrmion field in this coordinate system are
\begin{equation}
\label{Eqskyrmionline1}
\Sigma_u = \frac{1}{2}\varepsilon_{pqr}\ S_p\left(\frac{\partial S_q}{\partial v}\frac{\partial S_r}{\partial w}-\frac{\partial S_q}{\partial w}\frac{\partial S_r}{\partial v}\right) \, ,
\end{equation}
\begin{equation}
\label{Eqskyrmionline2}
\Sigma_v = \frac{1}{2}\varepsilon_{pqr}\ S_p\left(\frac{\partial S_q}{\partial w}\frac{\partial S_r}{\partial u}-\frac{\partial S_q}{\partial u}\frac{\partial S_r}{\partial w}\right) \, ,
\end{equation}
\begin{equation}
\label{Eqskyrmionline3}
\Sigma_w = \frac{1}{2}\varepsilon_{pqr}\ S_p\left(\frac{\partial S_q}{\partial u}\frac{\partial S_r}{\partial v}-\frac{\partial S_q}{\partial v}\frac{\partial S_r}{\partial u}\right) \, .
\end{equation}
The derivatives of the Stokes parameters are zero along the direction ${\bf u}$ and it follows, therefore, that 
$\Sigma_v=0=\Sigma_w$.  The one remaining non-zero component of the Skyrmion field at ${\bf r}_0$ is $\Sigma_u$
and it follows, therefore, that the Skyrmion field line at any point, ${\bf r}_0$, points along the direction of
constant polarisation.  Further details and consequences of this may be found in~\cite{Steve2023}. In a typical Skyrmionic optical beam with $n=1$ the lines of constant polarisation form a twisted mesh of straight lines shown in  {Fig.~\ref{fig:line} b.}

The identification of Skyrmion field lines with lines of constant polarisation is general and holds whether or not
the structured light has a non-zero Skyrmion number.  As such, Skyrmion field lines provide a natural way to extend
studies of L and C lines to arbitrary polarisations.  A fully general comparison, however, requires
an extension of the ideas presented here to non-paraxial fields.  We return to this point elsewhere.


\section{The Skyrmion vector potential and Dirac strings}

We have seen that the Skyrmion field is transverse and that this property allows for the introduction of a Skyrmion
vector potential ${\bf V}$.  As with the magnetic vector potential in electromagnetism, ${\bf V}$ is not unique and
may be modified by an analogue of a gauge transformation.  The Skyrmion number can be evaluated by performing a line
integral of the Skyrmion vector potential in much the same way that the magnetic flux through a surface is equivalent to 
an integral of the vector potential along a line bounding the surface.

A key feature in evaluating the Skyrmion number using the Skyrmion vector potential is the necessity of omitting
singularities by deforming the integration contour.  These singularities exist along lines in space and are analogous to
Dirac strings that appear in the theory of magnetic monopoles~\cite{Dirac,Frankel,Manton}.  Dirac strings are lines along
which the magnetic vector potential is singular.  They originate and end at magnetic monopoles or extend to infinity.  
For paraxial optical Skyrmions there is no analogue of the magnetic monopole and so the singular lines of the
Skyrmion vector potential are unbounded.

Dirac strings are not unique; a gauge transformation will move them.  The same behaviour holds for the Skyrmion 
vector potential and this means that there exists a variety of patterns of singular lines of the Skyrmion 
vector potential for any given Skyrmion field.  The contour required to evaluate the line integral of ${\bf V}$
will also vary, but the Skyrmion field and consequently the pattern of Skyrmion field lines will remain unchanged.
In the following section, we give an explicit example of this behaviour where we give two distinct forms of the Skyrmion vector potential for the same structures paraxial light beam and, consequently, the same Skyrmion field.

A pattern with Skyrmion number $n$ will typically be associated with $2n$ singular lines, or strings.  A special case 
exists, however, in which there is a single string, of strength $n$.  A continuous gauge transformation reveals, however,
further strings entering from infinity.  We shall provide an example of this behaviour in section \ref{SectOAM}.


\section{Skyrmions and optical orbital angular momentum}

\label{SectOAM}

The theory outlined above is rather general and facilitates the exploration of a very wide range of polarisation patterns. It is useful, however, to study a few simple examples to illustrate a few key features. To this end we select the two orthogonal spatial modes to be Laguerre-Gaussian modes of the form
\begin{equation}
\begin{split}
\label{Eq4.1}
u_p^\ell(\rho,\phi,z) &= \sqrt{\frac{2p!}{\pi(p+|\ell|)!}}\frac{1}{w(z)}\left(\frac{\rho\sqrt{2}}{w(z)}\right)^{|\ell|}
\exp\left(-\frac{\rho^2}{w^2(z)}\right) \\
& \times L_p^{|\ell|}\left(\frac{2\rho^2}{w^2(z)}\right)e^{i\ell\phi}
\exp\left(-i\frac{\rho^2}{w^2(z)}\frac{z-z_0}{z_R}\right) \\
& \times \exp\left(-i(2p + |\ell| + 1)\tan^{-1}\left(\frac{z - z_0}{z_R}\right)\right)
\end{split}
\end{equation}
where $z_R = \pi w_0^2/\lambda$ is the Rayleigh range, $w(z) = w_0\sqrt{1 + (z-z_0)^2/z_R^2}$ is the beam width at $z$ and $z=z_0$ is the
focal plane.  These have been much studied in the physics of lasers~\cite{Siegman,Milonni} and also by virtue of the orbital angular momentum,
of $\ell\hbar$ per photon, associated with them~\cite{Les,OAMbook,Alison,Bekshaev,Alison,Andrews}.

Selecting the modes $u_0({\bf r})$ and $u_1({\bf r})$ to be Laguerre-Gaussian modes means that the function $\mu$ reduces to the simple form
\begin{equation}
\label{Eq4.2}
\mu  =f(\rho ,z)e^{i\Phi (\rho ,\phi ,z)}=f(\rho ,z)e^{i\Theta (\rho ,z)}e^{i(\ell _1-\ell _0)\phi}
\end{equation}
so that the components of ${\bf S}$ are
\begin{align}
\label{Eq4.3}
S_x &= \frac{2f\cos[\Theta + (\ell_1 - \ell_0)\phi]}{1 + f^2} \, ,  \nonumber \\
S_y &= \frac{2f\sin[\Theta + (\ell_1 - \ell_0)\phi]}{1 + f^2} \, ,  \nonumber \\
S_z &= \frac{1-f^2}{1+f^2}  \, .
\end{align}
The symmetry of our light beam suggests that we express the components of {\boldmath$\Sigma$} in cylindrical polar coordinates~\cite{Parax}:
\begin{align}
\label{Eq4.4}
\Sigma_\rho &= -\frac{2(\ell_1 - \ell_0))}{\rho(1+f^2)^2}\frac{\partial f^2}{\partial z} \, ,  \nonumber \\
\Sigma_\phi &= -\frac{2}{(1+f^2)^2}\left(\frac{\partial f^2}{\partial \rho}\frac{\partial \Theta}{\partial z} - 
\frac{\partial f^2}{\partial z}\frac{\partial \Theta}{\partial \rho} \right) \, , \nonumber \\
\Sigma_z &= \frac{2(\ell_1 - \ell_0)}{\rho(1+f^2)^2}\frac{\partial f^2}{\partial \rho}  \, .
\end{align}
The Skyrmion number from Eq.~\ref{Eq3.9}. is then simply given by the integral of $\Sigma_z$ over a plane of constant $z$:
\begin{align}
\label{Eq4.5}
n(z) &= \frac{1}{4\pi}\int_0^\infty \rho\,\mathrm{d}\rho \int_0^{2\pi} \mathrm{d}\phi\,\Sigma_z \nonumber \\
&= (\ell_1 - \ell_0)\left(\frac{1}{1+f^2(0,z)} - \frac{1}{1+f^2(\infty,z)}\right)  \, .
\end{align}
It is remarkable that this depends only on the dominant polarisation on the $z$-axis where the optical vortex is situated and at large
distances from the axis.  If the mode $u_0({\bf r})$ dominates at $\rho = 0$, then $\lim_{\rho \rightarrow 0}\mu({\bf r}) = 0$ and
$f(0,z) = 0$, but if mode $u_1({\bf r})$ dominates then $f(0,z) = \infty$.  The value of $f(\infty,z)$ is similarly determined by the 
dominant spatial mode at large distances from the $z$-axis.

\begin{figure}[htbp]
    \centering
    \includegraphics[width=14cm]{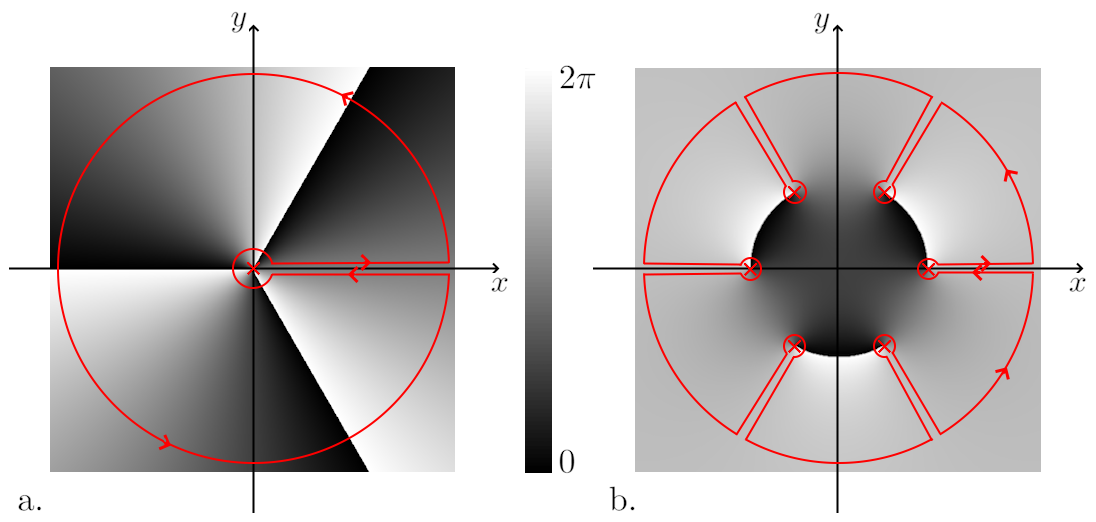}
    \caption{Integration contours for an $n=3$ skyrmion: (a) For $S_\pm=S_x\pm i S_y$. (b) For $S_\pm=S_z\pm i S_x$ (cf. Eq.~\ref{EqCS}).  The background shows the corresponding phase, $\Phi$, with phase singularities marked by red crosses.
    }
    \label{fig:exclusion_centre}
\end{figure}

We can shed some light on the form of our expression for the Skyrmion number by calculating it using the Skyrmion potential ${\bf V}$.
To do so we recall, from Eq. (\ref{Eq2.21}) that we can use the Skyrmion vector potential ${\bf V}$ to rewrite the Skyrmion number in the form
\begin{equation}
\label{Eq4.6}
n(z) = \frac{1}{4\pi}\oint {\bf V}\cdot \mathrm{d}\mbox{\boldmath$\ell$} \, .
\end{equation}
The integration contour must exclude singular points as showed in Fig.~\ref{fig:exclusion_centre} for two different choices of basis.  In the case of Fig.~\ref{fig:exclusion_centre}(a), where $S_{\pm}=S_x\pm iS_y$, this includes the $z$-axis where the argument of $\mu$ is not defined, and integrating around the contour and noting that the two straight line contributions to the contour cancel results in:
\begin{equation}
\label{Eq4.6a}
n(z) = \frac{1}{4\pi}\left(\lim_{\rho \rightarrow \infty}\int_0^{2\pi} \rho\,\mathrm{d}\phi\,\text{V}_\phi  - \lim_{\rho \rightarrow 0}\int_0^{2\pi} \rho\,\mathrm{d}\phi\,\text{V}_\phi\right) \, ,
\end{equation}
where 
\begin{equation}
\label{Eq4.6b}
\text{V}_\phi = \frac{S_z}{\rho} \frac{\partial}{\partial \phi}\arg(\mu) =  \frac{S_z}{\rho}(\ell_1 - \ell_0) \, .
\end{equation}
It follows that the Skyrmion number is
\begin{align}
\label{Eq4.7}
n(z) &= \frac{1}{2}(\ell_1 - \ell_0)\left(-\frac{1 - f^2(\infty,z)}{1 + f^2(\infty,z)} + \frac{1 - f^2(0,z)}{1 + f^2(0,z)}\right)  \nonumber \\
&= (\ell_1 - \ell_0)\left(\frac{1}{1 + f^2(0,z)} - \frac{1}{1 + f^2(\infty,z)}\right) \, .
\end{align}
The Skyrmion number will typically be $\pm(\ell_1 - \ell_0)$ or $0$, with the value determined by the modes that dominate at $\rho = 0$
and as $\rho \rightarrow \infty$.  It will be zero if a single mode dominates in both places, and non-zero if the polarisation at $\rho = 0$
is orthogonal to that for $\rho \rightarrow \infty$.  There is one exception, which is when neither mode is sufficiently dominant at $\rho = 0$ and in this case we find a non-integer Skyrmion number.  We shall deal with this case separately in the next
section. 

As noted above, the Skyrmion vector potential is not unique and we can change its form, without changing the Skyrmion
field, by a simple rotation of the basis.  This will not affect the Skyrmion field lines, and so will leave the 
Skyrmion number unchanged, but it will change ${\bf V}$ and also modify the pattern of singularities.  An example of this
is shown in Fig.~\ref{fig:exclusion_centre}(b), where we have rotated the Stokes vector through $-\pi/2$ about the
$y$-axis, so that $S_\pm$ becomes $S_z\pm iS_y$.  We see that the sole singularity at the origin in Fig.~\ref{fig:exclusion_centre}(a), around which the phase changes by $6\pi$ corresponding to a Skyrmion number of 3, is replaced by six, that is $2n$, 
strings around each of which the phase changes by $2\pi$. The Skyrmion number is, of course, unchanged.  We note
that such a change of basis can be advantageous from an experimental perspective as the integration contour can be moved to regions of higher intensity, yielding better measured Skyrmion numbers~\cite{mcwilliam}.

If our two spatial modes are a pair of Laguerre-Gaussians, of the form given in Eq.~(\ref{Eq4.1}) then the forms of $f(0,z)$ and of
$f(\infty,z)$ are particularly simple.  At large distances from the beam axis, the behaviour is determined solely by the rate at which 
the Gausssian tends to zero.  For large $\rho$ we find
\begin{equation}
\label{Eq4.8}
f(\rho,z) = \frac{P(\rho)}{Q(\rho)}\exp\left[-\rho^2\left(\frac{1}{w_1^2(z)} - \frac{1}{w^2_0(z)}\right)\right] \, ,
\end{equation}
where $P(\rho)$ and $Q(\rho)$ are polynomials.  As $\rho \rightarrow \infty$, $f(\rho)$ tends to zero if $w_1(z) < w_0(z)$ and if 
$w_1(z) > w_0(z)$ then $f(\rho)$ tends to infinity.  For small values of $\rho$, however, $f(\rho)$ is proportional to a power of
$\rho$ determined by the orbital angular momenta of the two modes:
\begin{equation}
\label{Eq4.9}
f(\rho) \propto \rho^{|\ell_1| - |\ell_0|} \,.
\end{equation}
This tends to zero if $|\ell_1| > |\ell_0|$ and to infinity if $|\ell_1| < |\ell_0|$.  Situations in which these quantities are equal are a special case 
that will be dealt with in the next section.  Note that the behaviour for small $\rho$ is quite general and applies for modes other than the 
Laguerre-Gaussians with a charge $\ell$ vortex.

The form of $\mu$ implicit in Eq. (\ref{Eq4.8}) reveals, also, how the Skyrmion number can change on propagation.  We have
seen how the widths $w_0(z)$ and $w_1(z)$, when combined with the behaviour at the origin, determine the Skyrmion number.
If the two Laguerre-Gaussian modes are focused at different points, then the relative widths of the modes can switch, say 
from $w_1(z)<w_0(z)$ to $w_1(z)>w_0(z)$ and with this the form of the polarisation at large distances, with the 
corresponding change in the Skyrmion number~\cite{Parax}.  When this happens, Skyrmion field lines turn away from the
propagation direction heading radially outwards.


\section{Non-integer Skyrmions}
\label{Sect.fractional}

The remaining case to consider is when the two spatial modes have equal but opposite orbital angular momentum so that
$\ell_1 = -\ell_0 = \ell$. This case can also produce a Skyrmion structure but this time the Skyrmion number takes a fractional, or
more precisely a non-integer, value.  To the best of our knowledge, this situation does not appear in other fields of physics in which 
Skyrmions or the associated textures arise.

We shall see that non-integer Skyrmions have a number of features that distinguish them 
from the more familiar integer Skyrmions described above.

To begin our discussion let us return to our formula for the Skyrmion number associated with a pair of spatial modes carrying orbital 
angular momentum, Eq.~(\ref{Eq4.5}).  The Skyrmion number depends only on the function $f(\rho,z)$ at large and small values of 
$\rho$.  For large $\rho$ one of the two modes will always dominate and hence if we move far enough from the $z$-axis we will find 
that $f(\rho,z)$ tends to zero or to infinity.  The new feature occurs on the $z$-axis where as $\rho \rightarrow 0$ we find that 
$f(\rho,z)$ tends to the value
\begin{equation}
\label{Eq5.1}
f(0,z) = \sqrt{\frac{p_1!(2p_0 + |\ell|)!}{p_0!(2p_1+|\ell|)!}} \left|\frac{\beta}{\alpha}\right|\left(\frac{w_0(z)}{w_1(z)}\right)^{|\ell|+1} \, .
\end{equation}
It follows that the associated Skyrmion number,
\begin{equation}
\label{Eq5.2}
n(z) = 2\ell\left(\frac{1}{1+f^2(0,z)} - \frac{1}{1+f^2(\infty,z)}\right) \, ,
\end{equation}
can take any desired value.  As the beam propagates, moreover, this Skyrmion number will change as diffraction causes the 
ratio of beam waists $w_0(z)/w_1(z)$ to change.

The fact that the Skyrmion number changes on propagation requires an explanation.  The situation is clearly distinct from that
produced by a pair of spatial modes focused in different places, as there the Skyrmion number changes abruptly at a single
plane.  This suggests that a different explanation is required.  To find this we note that the issue seems to originate from the
behaviour of the Skyrmion field on the $z$-axis and it is there that we find an explanation.  For small values of $\rho$ the ratio
of the mode amplitudes tends to the value
\begin{equation}
\label{Eq5.3}
\lim_{\rho \rightarrow 0}\mu = f(0,z)\exp\left(-i2(p_1 - p_0)\tan^{-1}\left(\frac{z - z_0}{z}\right)\right)e^{i2\ell\phi} \, .
\end{equation}\
As the magnitude of this is non-zero it is clearly ill-defined as the azimuthal angle $\phi$ has no meaning on the axis. 
It follows that the point $z=0$ in any given plane is singular.

\begin{figure}[htbp]
    \centering
    \includegraphics[width=4cm]{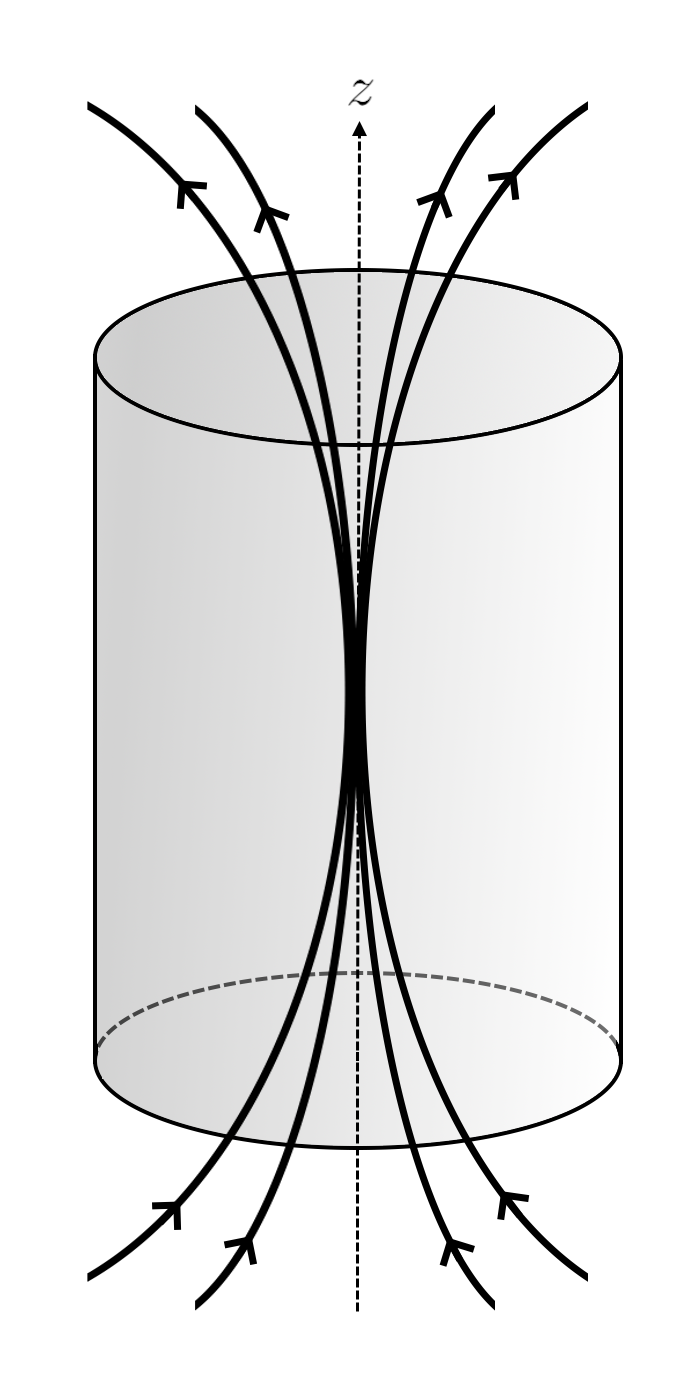}
    \caption{Lines of constant polarisation going in and out of the cylindrical volume for a non-integer Skyrmion. 
    }
    \label{fig:line2}
\end{figure}

To focus attention on the Skyrmion field on the $z$-axis, let us consider a cylindrical volume of radius $\Delta\rho$ and height $\Delta z$
centred on and rotationally symmetric about the $z$-axis, as depicted in Fig.~\ref{fig:line2}.  We can use our expressions for the $z$- and $\rho$-components of {\boldmath{$\Sigma$}}, given in Eq.~(\ref{Eq3.5}),
to evaluate the flux of the Skyrmion field through the surface of this cylinder.  We find
\begin{equation}
\label{Eq5.4}
\oint \mbox{\boldmath$\Sigma$}\cdot \mathrm{d}{\bf S} = 4\pi 2\ell\left(\frac{1}{1+f^2(0,z+\Delta z)} - \frac{1}{1+f^2(0,z)}\right) \, .
\end{equation}
This depends, clearly, only on the polarisation on the $z$-axis, through the function $f(0,z)$.  Gauss's theorem then implies that the 
divergence of {\boldmath{$\Sigma$}} is not zero but rather takes a singular value on the $z$-axis,
\begin{equation}
\label{Eq5.5}
\mbox{\boldmath$\nabla$}\cdot\mbox{\boldmath$\Sigma$} = 4\ell\frac{\delta(\rho)}{\rho}\frac{\partial}{\partial z}\frac{1}{1+f^2(\rho,z)} \, ,
\end{equation}
so that the volume integral of this over our cylinder gives
\begin{align}
\label{Eq5.6}
\int \mbox{\boldmath$\nabla$}\cdot\mbox{\boldmath$\Sigma$}\,\mathrm{d}\Omega &= 8\pi \ell\left(\frac{1}{1+f^2(0,z+\Delta z)} - \frac{1}{1+f^2(0,z)}\right) 
\nonumber \\
&= \oint \mbox{\boldmath$\Sigma$}\cdot\mathrm{d}{\bf S} \, ,
\end{align}
as required.  The theorem, which we proved earlier (\ref{Eq2.4}), established that $\mbox{\boldmath$\nabla$}\cdot\mbox{\boldmath$\Sigma$} = 0$,
but here we see the need to add a caveat to this due to the singular behaviour of the Skyrmion field on the $z$-axis.
For a non-integer Skyrmion, the Skyrmion field does not have a properly defined value on the $z$-axis and therefore does not have a well-defined derivative there. 
Our earlier proof of the divergenceless nature of {\boldmath$\Sigma$} employed the idea that 
${\bf S}$ (or ${\bf M}$) is slowly varying in that its value can be represented by a Taylor series in the vicinity of any given point.  The 
singular form of the Skyrmion field on the $z$-axis means that the required slowly varying behaviour does not hold on the $z$-axis.
The fact that $\mbox{\boldmath$\nabla$}\cdot\mbox{\boldmath$\Sigma$} = 0$ everywhere {\it except} on the $z$-axis means that 
the Skyrmion field lines can originate or terminate there, and it is by this mechanism that the Skyrmion number can change continuously 
as the light propagates.  A schematic illustration of this behaviour is presented in Fig.~\ref{fig:line2}. The failure of the transversality condition for non-integer Skyrmions also suggests that
a description of these in terms of an underlying Skyrmion vector potential may not be helpful.

\section{Conclusion}
While optical Skyrmions fall into a specific category of structured light beams, studying their fundamental properties uncovers powerful tools for examining the topological properties of paraxial light beams. Optical Skyrmions are 
best understood by reference to a Skyrmion field. This field is transverse, or divergenceless, and so may be represented
in terms of a vector potential.
This potential, although not unique, provides a simple expression for the Skyrmion number in paraxial optical Skyrmion fields based solely on topological considerations. 
Skyrmion field lines have a significance beyond their association with Skyrmions: they are lines of constant 
polarisation~\cite{Steve2023}.  As such they underlie and map the forms of all structured paraxial beams.  It is likely,
therefore, that they will find general application in the developing study of structured light beams~\cite{Forbes21,Rubinsztein-Dunlop}.
Extending the study of Skyrmions beyond the paraxial regime could bring novel insights into topological charge conservation as has been advocated in the study of L and C lines~\cite{Nyebook,Nye,Berry,Dennis_Pi}.

It is straightforward to construct paraxial beams with Skyrmionic structure by superposing orthogonally polarised Laguerre-Gaussian modes.  For these we can determine the Skyrmion number purely in terms of the polarisation on
the beam axis and at large distances.  We can readily construct analogues of the Skyrmions encountered in the 
study of magnetic media.  The spins in a magnetic film interact, but there is no such interaction for
a light beam.  For this reason, we can contemplate creating structures that do not exist in magnetism, such as Skyrmions with arbitrary polarisation at their centre. A further example given here is the existence of non-integer Skyrmions.  These have many of the features of the more familiar 
integer Skyrmions but lack their stability in that the Skyrmion number typically varies on propagation.  The origin of 
this effect is the failure of the transversality or divergenceless condition on the beam axis.  It is this feature
that accounts for both the fractional (or more precisely non-integer) nature of the Skyrmions and for their 
changes on propagation.\vskip6pt

\enlargethispage{20pt}

\section{Acknowledgements}

\noindent S. M. B. acknowledges support from the Royal Society, grant number RP150122, J. B. G acknowledges support from the EP/V048449/1 grant and the Leverhulme Trust, A. M. acknowledges support by a UK Research and Innovation Council grant EPSRC/DTP 2020/21/EP/T51786/1. C. M. C. acknowledges financial support from UK Research and Innovation Council (UKRI) grant EP/W016486/1.


\end{document}